\documentclass[letter]{aa}

\usepackage{graphicx}
\usepackage{txfonts}
\usepackage{multirow}
\usepackage{natbib}
\usepackage{color}
\usepackage{amsmath}
\newcommand{\masyr}{${\rm mas\, yr^{-1}}$}

\newcommand{\xsun}{X_\odot}
\newcommand{\ysun}{Y_\odot}
\newcommand{\zsun}{Z_\odot}
\newcommand{\vxsun}{V_{X,\odot}}
\newcommand{\vysun}{V_{Y,\odot}}
\newcommand{\vzsun}{V_{Z,\odot}}
\newcommand{\Msun}{M_\odot}
\newcommand{\gaia}{{\it Gaia~}}

\begin{document}

   \title{The power of teaming up HST and {\it Gaia}: the first proper motion measurement of the distant cluster NGC~2419}

   \subtitle{}

   \author{D. Massari\inst{1,2}
          \and
          L. Posti\inst{2}
          \and
          A. Helmi\inst{2}
          \and
          G. Fiorentino\inst{1}
          \and
          E. Tolstoy\inst{2}
          }

   \institute{INAF-Osservatorio Astronomico di Bologna, via Ranzani 1, 40127, Bologna, Italy\\
              \email{davide.massari@oabo.inaf.it}
         \and
            University of Groningen, Kapteyn Astronomical Institute, NL-9747 AD Groningen, Netherlands
             }

   \date{Received XXX; accepted YYY}

% \abstract{}{}{}{}{}
% 5 {} token are mandatory

  \abstract
   {}
   {We present the first measurement of the proper motion and orbit of the very distant and intriguing globular cluster NCG~2419. }
   {We have combined data from HST and Gaia DR1 to derive the relative
     proper motions of stars in the direction to the cluster. To tie
     to an absolute reference frame we have used a background galaxy located in the
     field.}
   {We find the absolute proper motion of NGC~2419 to be
     $(\mu_{\alpha}\cos(\delta)$, $\mu_{\delta}$)=($-0.17\pm0.26,
     -0.49\pm0.17$)~\masyr. We have integrated the orbit of the
     cluster in a Galactic potential and found it to oscillate between
     $\sim$53 kpc and $\sim$98 kpc on a nearly polar orbit.  This
     makes it very likely that NGC~2419 is a former cluster of the
     Sagittarius dwarf spheroidal galaxy, also because it shares the
     same sense of rotation around the Milky Way.
    %This new
    % measurement will undoubtedly provide new constraints on both the shape as
    % well as radial density profile of the dark matter halo of our
    % Galaxy.
    }
   {}

   \keywords{astrometry – proper motions – globular clusters: individual: NGC~2419}
   
   \maketitle
%
%________________________________________________________________

\section{Introduction}

Proper motions (PMs) are an extremely powerful tool to investigate the
evolution of the Milky Way and its satellites.  In the context of a
hierarchical growth of structure, many of the objects \citep[stars,
streams, globular clusters, e.g.][]{helmi16,bel06,dinescu2002} that we
currently observe in the Milky Way halo could have originated outside
the Galaxy, brought in by satellites that have survived until the
present day or accreted a long time ago. Typically photometric
information alone is not enough to establish such origins
robustly. Possibly the most effective way is to determine the orbital
trajectories and this requires their PMs to be measured.  However,
since the PMs size depends on the distance and on the temporal
baseline between the astrometric observations, PMs can be very small
and difficult to measure.

Nevertheless, a new golden era for PMs science is rising.  The {\it Hubble
  Space Telescope} has been the most powerful astrometric instrument
so far. Thanks to its Point Spread Function (PSF) and geometric
distortions being stable within a few percent (\citealt{jayacs, bellini11}),
HST has reached astrometric accuracies on single exposures up to
$\sim0.5$ mas for bright stars (\citealt{jayacs}). Very recently, the
astrometric satellite \gaia released its first positional measurements
(Data Release 1, \citealt[][Brown et al.]{brown16}) for more than 1 billion stars
across the whole sky, and a sub-sample of these have positional
accuracies comparable to HST.
However, these two exceptional
instruments have limitations if taken separately.  For example, HST
often lacks second epoch observations, while at the end of its
mission, \gaia will provide PMs obtained on a temporal baseline of only
five years. Yet the synergy between HST and \gaia allows us 
to overcome the main limitations of both: \gaia will provide accurate second epoch measurements for
any HST data, while in several cases, especially for the central regions of crowded stellar systems,
HST will provide observations taken up to
$15$-$20$ years before {\it Gaia}, thus increasing the PMs temporal baseline
by a factor of $4$-$5$.

In this work we describe the first combination of datasets
from HST and \gaia to measure the PM and determine the orbit of the
enigmatic cluster NGC~2419. NGC~2419 is by far the brightest globular
cluster in the outer regions of the Milky Way halo.  Its chemistry is
comparable to that of massive clusters, with homogeneous [Fe/H]
measurements (\citealt{mucciarelli12}), and photometric evidence for He
enhancement (\citealt{dicri11a}). On the other hand, there are
suggestions of a Calcium spread (\citealt{cohen10, lee13}), and its
half-light radius is significantly larger than that of any typical
globular cluster (\citealt{harris}).  This, coupled with its unusually
large distance (87.5 kpc, \citealt{dicri11b}), makes it
intriguing. NGC~2419 has been proposed as a candidate
nucleus of an accreted dwarf galaxy, as well as for having been formed
inside a dark matter halo (\citealt{baumgardt09}). 
To shed light on its nature and origin, we
present the first estimate of the proper motion of NGC~2419, by exploiting 
the combination of HST and {\it Gaia}. This is then used to compute possible 
orbits in a realistic Galactic mass distribution.
An analysis that is similar in spirit has been performed on five nearby globular clusters by
\cite{watkins16} who exploited the Tycho-Gaia solution (TGAS, see \citealt{lindegren2016}) rather than using HST.
However, since TGAS PMs are available only for bright stars, no such measurements exist for 
NGC~2419 members.

In Section \ref{method} we describe the data analysis and the method
we used to measure the absolute PM of the cluster. In Section
\ref{orbit} we compute the orbit of NGC~2419 and check for possible
associations with dwarf galaxies and other known stellar
clusters. We summarize our conclusions in Section \ref{concl}.

\section{Data analysis and PM measurement}\label{method}

As the first epoch for the PM determination we used data acquired under
GO-9666 (PI:R. Gilliland) with the Wide Field Channel (WFC) of the
Advanced Camera for Survey (ACS) on board the HST. The WFC/ACS is made
up of two detectors with size $2048 \times 4096$ pixel and a pixel
scale of $\sim0.05 \arcsec\,$pixel$^{-1}$, which are separated by a
gap of about 50 pixels, so that the total field of view (FoV) is
$\sim200\arcsec \times 200\arcsec$.  We used
$14$ deep exposures in the F435W, F475W, F555W, F606W, F625W,
F775W, F814W filters (2 exposures per filter), taken on September 26,
2002.  We work with \_FLC images, which have been corrected by the
HST calibration pipeline for charge transfer efficiency
(\citealt{jaybedin10} and \citealt{ubedajay}).  The
data-reduction is based on the procedures described in
\cite{jayking06}. Each individual exposure was analysed with the
publicly available program \texttt{img2xym$\_$WFC.09$\times$10}.  This
program uses a pre-determined model of the PSF plus a single
time-dependent perturbation, and produces a catalogue with
positions and instrumental magnitudes as output.  
After rejecting all the saturated sources, the stellar positions 
in each catalogue were corrected for filter-dependent geometric 
distortions, using the solution provided by \cite{jayacs}.

The second epoch data are provided by the \gaia Data Release 1 (DR1, see \citealt[][Brown et al.]{brown16}).
DR1 positions are from January 1, 2015. In combination with HST, this provides a temporal
baseline for the PM measurement of $12.27$ years.
We requested from the \gaia archive ({\tt https://gea.esac.esa.int/archive/}) a catalogue containing
positions, related uncertainties, $G$ magnitudes and astrometric excess noise for all the
sources in the FoV covered by the HST dataset.
We found that the median positional error for this catalogue was $\sim0.6$~mas, 
and we decided to exclude from the analysis all the sources with
a positional error larger than $3$~mas ($\sim5$ times the median error value) to remove poorer quality 
measurements. 

The PMs were measured using the procedure described in \cite{massari13}. We chose as master frame
that described by the \gaia positions, which is already aligned with the equatorial
coordinate system. Then we transformed each HST single exposure catalogue onto the master
frame using a six-parameter linear transformation. To maximize the accuracy of these
transformations we treated each chip of the HST exposures separately to avoid spurious
effects due to the presence of gaps.
After this process, each source had up to 14 first-epoch positions transformed onto the
master frame. We decided to exclude from the following analysis all those sources with less than
$4$ first-epoch detections. The PMs of the remaining 481 objects were computed as the difference between the second epoch \gaia
positions and the 3$\sigma$-clipped median value of the HST first-epoch positions, divided the temporal
baseline. The two projected components of the PMs on the sky were treated separately.
The uncertainties on the PMs were computed as the sum in quadrature between the \gaia positional errors
and the rms of the residuals about the median value of HST positions, divided by the temporal baseline.

\begin{figure}
    \includegraphics[width=\columnwidth]{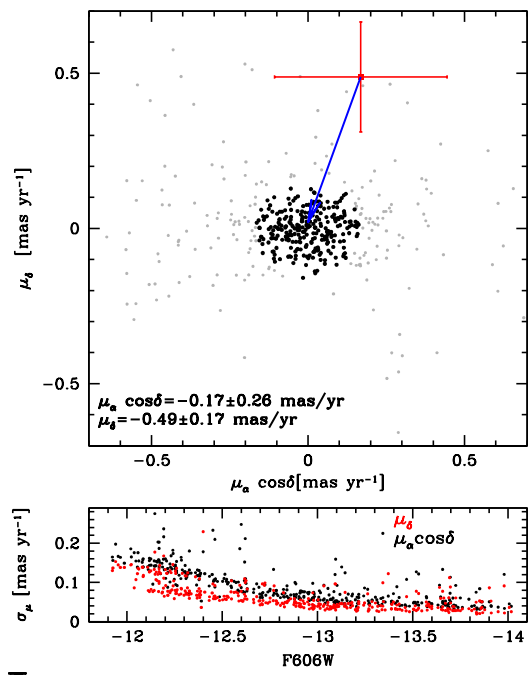}
        \caption{\small {\it Top panel:} VPD for the stars in our final catalogue. Stars used to measure the average
        cluster PM are highlighted in black, whereas likely non-members are shown in grey. The location of the background galaxy
        used to determine the absolute PM zero-point with its uncertainty is shown with a red symbol. {\it Bottom panel:} Uncertainties on the PM
        measurements. Black and red symbols are related to the two different PM components as described in the labels.}\label{vpd}
\end{figure}

After this first iteration, we repeated the procedure by computing the
frame transformations using only likely cluster members. We selected
stars according to both their location in the colour-magnitude diagram
(CMD) and their first PM determination, requiring consistency with the
mean cluster motion (which by construction is centred on
[0,0]~\masyr).  After four iterative steps, the selected number of
stars ceased to change ($366$ were used in the last step), and the
resulting PMs and related errors are those of our final
catalogue. The PMs are shown in the top panel of Fig.\ref{vpd}, also
known as a Vector Point Diagram (VPD), while the uncertainties for
each PM component are plotted in the bottom panel with different colours.
The larger uncertainties on $\mu_{\alpha}\cos(\delta)$ are due 
to the positional errors in the Gaia dataset, and explain why the distribution 
of stars in the VPD appear elongated in that direction.

\begin{figure}
    \includegraphics[width=\columnwidth]{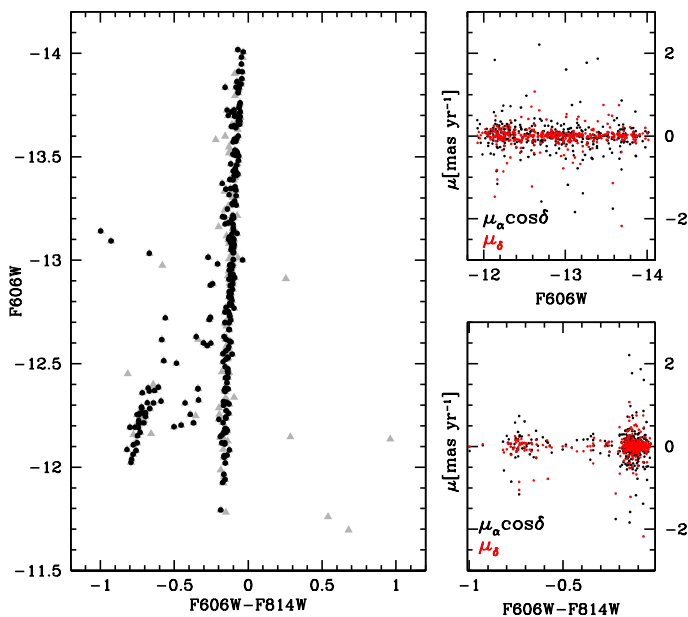}
        \caption{\small {\it Left panel:} (F606W, F606W-F814W) instrumental CMD of NGC~2419 for the stars in the final PM catalogue.
        Black and grey symbols indicate stars selected in the VPD as shown in Fig.\ref{vpd}. {\it Right panels:} behaviour of the two PM
        components with respect to instrumental magnitude (top panel) and colour (bottom panel). In both cases, the best fits
        indicate no systematic trend.}\label{cmd}
\end{figure}

We have performed several consistency checks on these PM measurements.
First, we verified that the bulk of the PM distribution centred
around zero is actually made up of cluster member stars. We selected
stars around the mean PM value in the VPD with an iterative
2.5$\sigma$-clipping procedure. Their location in the instrumental
(F606W, F606W-F814W) CMD is shown with black symbols in the left panel
of Fig.~\ref{cmd}.  All the selected stars lie on the cluster
evolutionary sequence, and other stars that are also on this sequence
(grey symbols) are excluded because of their large PM uncertainties
(see Fig.~\ref{vpd}). Following \cite{bellini14}, \cite{massari16}, we also
checked for spurious systematic trends of the measured PM components
with spatial distribution, instrumental magnitude and colour, and
found none. This is demonstrated in the top- and bottom- right panels
of Fig.~\ref{cmd}. The distributions with magnitude and colour are
consistent with no systematic trends within a 1$\sigma$ uncertainty.
All these checks support the quality and the reliability of
our measurements. However, since we are using only two epochs, we cannot
exclude that other subtle systematic errors affect our analysis, possibly making
the overall estimate of the PM uncertainties a lower limit.

\begin{figure}
    \includegraphics[width=\columnwidth]{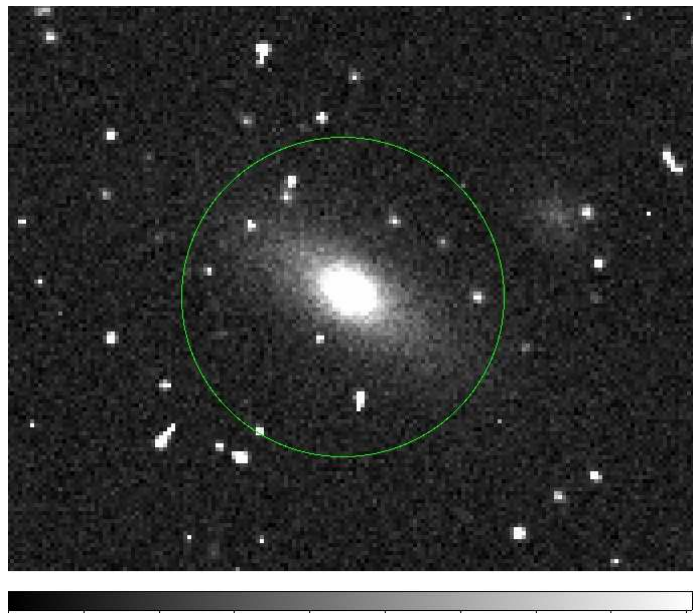}
        \caption{\small Background galaxy in one HST F606W exposure used to determine the absolute PM zero points.}\label{gal}
\end{figure}

The PMs measured in this way are relative to the mean motion of the
cluster. Relating them to an absolute reference frame is therefore
necessary in order to obtain the systemic motion of NGC~2419.  As
extensively demonstrated in the literature (e.g. \citealt{casetti04, massari13, sohn13, pryor15}) very distant
objects like quasars or background galaxies can be used
to determine the absolute PM zero point, as
their absolute PM is $\sim0$~\masyr. A search
through the NED revealed no such objects in our FoV. However when we
inspected the images by eye looking for background galaxies, we found
one object in both our HST and \gaia (which are
most shallow) data, that has an extended
structure typical of background galaxies as shown in
Fig.~\ref{gal}. This is confirmed by the corresponding {\it Gaia}
astrometric noise excess value which is larger than $10$. On the other
hand, its overall profile is point-like enough to be well described by
the adopted PSF as can be inferred from its QFIT value $<0.5$ (see
\citealt{jayking06}). Morever, it appears to be bright, with a signal-to-noise ratio
$>150$ in all the HST exposures, and isolated, with no neighbouring sources
affecting its centroid determination. This object thus has all the features required
to provide a reliable determination of the absolute PM zero point. Its location
in the Vector Point Diagram (VPD) with the corresponding uncertainties
is shown in Fig.\ref{vpd} in red, and is centred on
($\mu_{\alpha}\cos(\delta)$, $\mu_{\delta}$)=($0.17\pm0.26,
0.49\pm0.17$)~\masyr. Since the average PM of likely member stars
(black points) is ($\mu_{\alpha}\cos(\delta)$,
$\mu_{\delta}$)=($0.001\pm0.007, -0.004\pm0.006$)~\masyr, i.e. is
consistent with zero within $1\sigma$, the absolute PM of NGC~2419
is ($\mu_{\alpha}\cos(\delta)$,
$\mu_{\delta}$)=($-0.17\pm0.26, -0.49\pm0.17$)~\masyr.  In Galactic
coordinates, this corresponds to ($\mu_l \cos b, \mu_b$) = ($0.43\pm0.09,
-0.29\pm0.30$)~\masyr.

Given the large distance of NGC~2419,
systematic uncertainties due to global systemic motions of the cluster such
as expansion/contraction or rotation in the plane of the sky (see
\citealt{massari13}) are negligible compared to the uncertainty on the
absolute zero-point. For example, \cite{baumgardt09} found a rotation velocity
of 3.1 km s$^{-1}$ for NGC~2419 (with an r.m.s of
$4.0$ km s$^{-1}$), which translates into an additional systematic
error of only $0.007$ mas yr$^{-1}$.  

\section{The orbit of NGC~2419}\label{orbit}

By combining the above measurements with the cluster's radial velocity
(v$_{rad}=-20.3\pm0.7$ km s$^{-1}$, \citealt{baumgardt09}), distance
($87.5\pm3.3$ kpc, \citealt{dicri11b}) and sky position
(RA,Dec)=($114.535\pm0.004$, $+38.8824\pm0.0003$)~deg 
\citep{goldsbury10}, we are able to determine for the first time the orbit of NGC~2419. 

To this end, we transform these measurements to a heliocentric
right-handed Cartesian reference frame, where $X$ points towards the
Galactic centre, $Y$ in the direction of rotation and Z towards the Galactic North pole. 
This yields $(X,Y,Z) = (-79.1, -0.5, 37.4)$ in kpc, and $(V_X,V_Y,V_Z)=(-32.6,
-177.2, -119.3)$ in km s$^{-1}$.  We then transform to a Galactocentric reference
frame by assuming the the Sun's position and velocity to be
$(\xsun,\ysun,\zsun) = (-8.3, 0, 0.014)$~kpc, and
$(\vxsun,\vysun,\vzsun) = (11.1, 240.24, 7.25)$~km s$^{-1}$ 
\citep[see][]{Schonrich+10}.

We thus compute the orbit of NGC~2419 in a Galactic potential consisting of
a flattened bulge, a gaseous exponential disc, thin and thick stellar
exponential discs and a flattened dark matter halo \citep[for more
details, see][]{Piffl+14}.  The model has a total baryonic (stars and
cold gas) mass of $M_{\rm bary} = 5.3\times 10^{10}\Msun$ and a virial
halo mass of $M_{200}=1.3\times 10^{12}\Msun$.  The dark halo follows
the \citet[][]{NFW96} form, its flattening is 
$q=0.8$ and has a concentration of $c_{200}\simeq 20$.

We use the phase-space position of the cluster we have just derived
as the initial condition to integrate forward and
backward in time an orbit for about 4 Gyr using an 8$^{th}$ order
Runge-Kutta method. We also generate 100 realizations of the initial
phase-space coordinates by assuming that the errors in the space of
observables are Gaussian, and integrate them in the same way.

\begin{figure}
\includegraphics[width=\columnwidth]{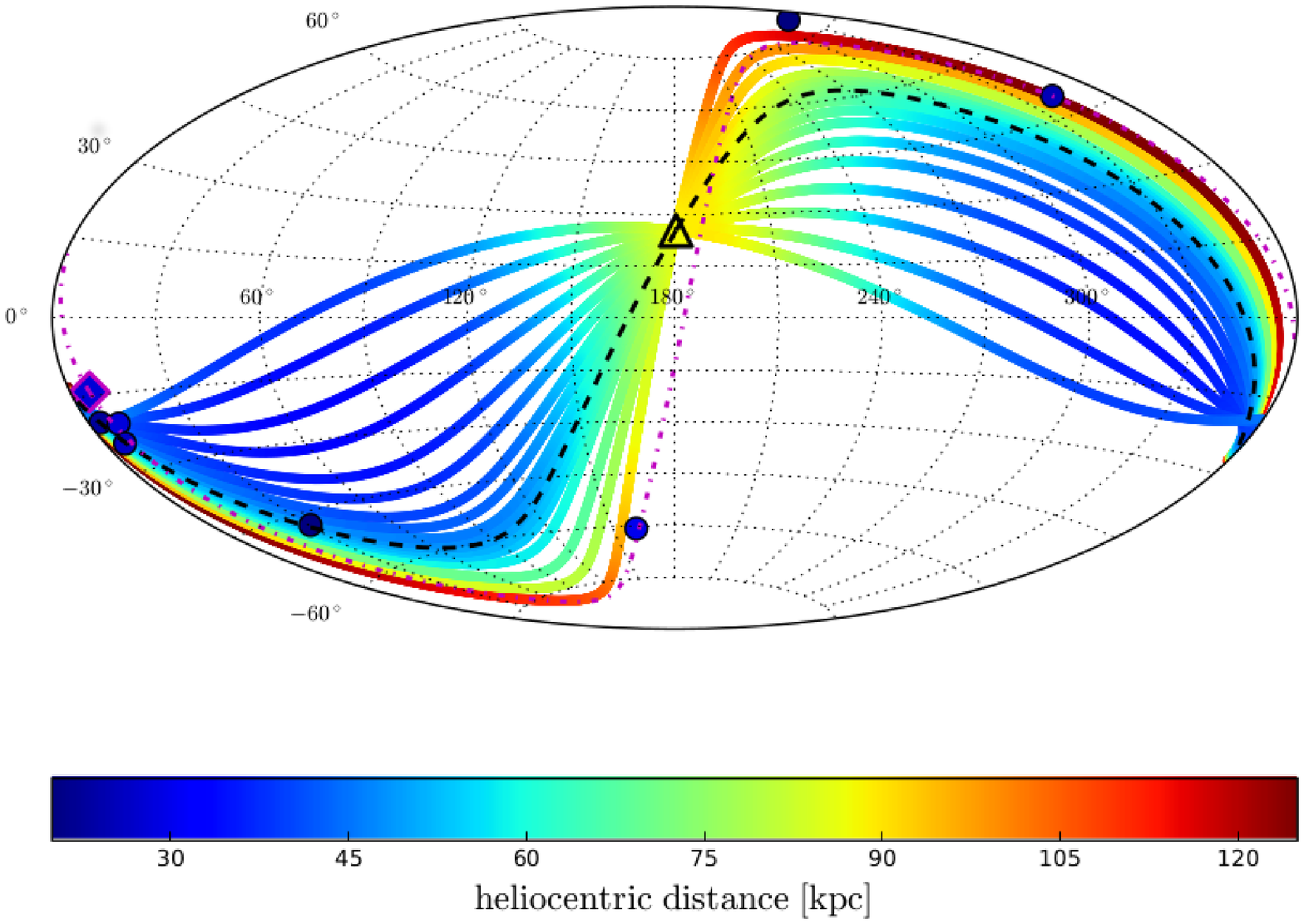} \\
\includegraphics[width=\columnwidth]{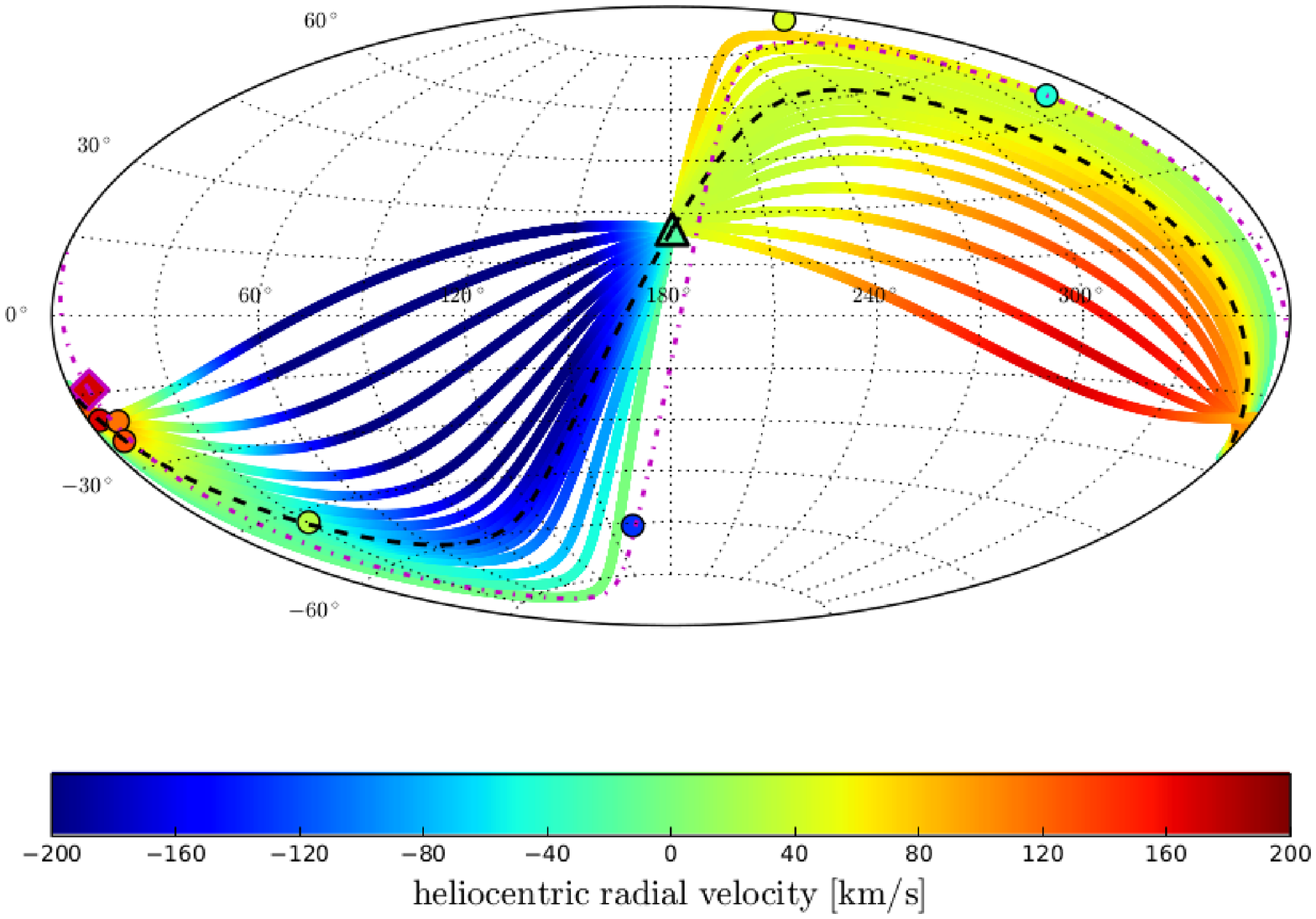}
\caption{\small{\it Top panel:} Trajectory on the sky of NGC~2419's possible
  orbits (within 1$\sigma$), where that obtained by starting from its
  current position and velocity (indicated by a triangle) is shown
  with the black dashed line.  The colour-coding represents the
  (heliocentric) distance. For comparison, we also show the current
  position and orbit of the Sagittarius dwarf spheroidal galaxy
  (diamond and magenta dot-dashed line), integrated in the same
  Galactic potential, and the positions of globular clusters (coloured
  circles) possibly associated to Sagittarius according to
  \cite{LawMajewski10b}.  {\it Bottom panel:} same as the top panel,
  but where the colour-coding represents the (heliocentric) radial
  velocity.}\label{fig:traj}
\end{figure}

Fig.~\ref{fig:traj} shows the trajectories on the sky of a subset of
the orbits obtained in this way, i.e. those with PM initial conditions
within 1$\sigma$ from the measured values. The large errors on the
measured PMs result in trajectories that cover a large portion of the
sky.  However more probable orbits typically do not deviate by more
than a few degrees from the mean orbit shown by the dashed
curve. Fig.~\ref{fig:traj} also shows that this orbit is close to
polar, indicating that most of the angular momentum is in the
Y-direction. We find the orbit rotates in the clockwise direction and
has pericentre and apocentre distances $r_{\rm peri} = 53^{+23}_{-26}$
kpc, and $r_{\rm apo} = 98^{+2}_{-1}$~kpc respectively.

Based on the position of NGC~2419 on the sky, \cite{Irwin99} suggested
that it may have been a globular cluster associated with the
dwarf spheroidal galaxy Sagittarius, that was lost as soon as this entered the
potential well of the Milky Way. Our PM measurements show
that the sense of
the rotation of Sagittarius and NGC2419 about the Galactic centre are
the same, and therefore an association appears rather likely.
Despite the fact that NGC~2419 lies at a much larger distance than
the current orbit of Sagittarius \citep[see e.g.][and the magenta line in Fig.\ref{fig:traj}]{LawMajewski10a},
it must be borne in mind that if Sagittarius was much more massive
in the past, its debris will be located at large range of distances
reflecting the initial energy spread \citep[e.g.]
[predicted debris to lie at distances close to 100 kpc; see also
\citeauthor{gibbons2014} \citeyear{gibbons2014}]{helmi2001}. 
Dynamical friction can also act in the same sense and make the orbit
of Sagittarius sink towards the Galactic centre with time.

Furthermore, \citet[][]{bel2014} recently suggested that a tidal
stream consisting of blue horizontal branch stars \citep[reported first
by][]{newberg2003} found to overlap with NGC~2419 spatially as well as
in line-of-sight velocity, is part of the trailing stream of
Sagittarius.  Although the original model of
\citet[][]{LawMajewski10a} predicted for the streams a different trend
of line-of-sight velocity with angular phase than observed, the more
recent model of \citet{vera-ciro2013} in which the dark halo of the
Galaxy is oblate near the centre, and significantly triaxial at large
distances (and which includes the gravitational effect of the Large
Magellanic Could), fares better. Interestingly,
the tangential velocity predicted by this model \citep[see the
right-hand-side of Fig.~5 in][]{vera-ciro2013}  is also in good agreement with what
we have just derived for NGC~2419. 

%% 180.37\deg, 25.24\deg,
%% 87.5\,km s$^{-1}$, 
%% mu_l cos b = 0.43{\rm mas\, yr^{-1}}, 
%% mu_b = -0.29{\rm mas\, yr^{-1}}

\section{Conclusions}\label{concl}

We have presented the first measurement of the proper motion of the
intriguing globular cluster NGC~2419 thanks to  the unique combination of HST and
\gaia data. By using a background galaxy to tie our measurements to an
absolute reference frame, we determined the absolute PM of NGC~2419 to
be ($\mu_{\alpha}\cos(\delta)$, $\mu_{\delta}$)=($-0.17\pm0.26,
-0.49\pm0.17$)~\masyr. 

Numerical integration of the possible orbits in a Galactic potential starting from
the current location and velocity of NGC~2419 show the cluster to be
on an elongated orbit with a pericentre at $\sim 53$~kpc and an apocentre at
$\sim 98$~kpc. Its orbit is close to polar and rotates in the same
sense around the Milky Way as the Sagittarius dwarf galaxy. Our
analysis suggests that it is very likely that NGC~2419 originated in
the Sagittarius system. By combining all the information we have about
Sagittarius, its streams and its likely former globular cluster
NGC~2419, we may also be very close to pinning down the gravitational
potential of the Milky Way at large radii, as well as reconstructing
the remarkable history of the Sagittarius dwarf galaxy.

\begin{acknowledgements}
  We thank the anonymous referee for comments and suggestions 
  which improved the presentation of our results.
  DM and GF have been supported by the FIRB 2013 (MIUR grant RBFR13J716).
  AH and LP acknowlegde financial support from a Vici grant from NWO.
  This work has made use of data from the European Space Agency (ESA)
  mission \gaia (http://www.cosmos.esa.int/gaia), processed by the
  \gaia Data Processing and Analysis Consortium (DPAC, {\tt
    http://www.cosmos.esa.int/web/gaia/dpac/consortium}). Funding for
  the DPAC has been provided by national institutions, in particular
  the institutions participating in the \gaia Multilateral Agreement.
\end{acknowledgements}

\bibliographystyle{aa} % style aa.bst
%   \bibliography{ms.bib} % your references Yourfile.bib
%\bibliography{ms.bib}

\end{document}